\documentstyle[prd,aps,epsf,preprint]{revtex}

\def\@pnumwidth{2em}
\makeatother

\begin{document}

\title{Ultra-High Energy Gamma Rays in Geomagnetic Field and Atmosphere}

\author{H.P.~Vankov$^{1}$, N.~Inoue$^{2}$ and K.~Shinozaki$^{3}$}

\address{{\rm
$^{1}$Institute for Nuclear Research and Nuclear Energy, Sofia 1784,
Bulgaria \\
$^{2}$Department of Physics, Saitama University, Saitama 338-8570,
Japan\\
$^{3}$Institute for Cosmic Ray Research, University of Tokyo, Chiba
277-8582, Japan\\}}

\date{\today}

\maketitle

\begin{abstract}
The nature and origin of ultra-high energy (UHE: referring to $>10^{19}$
eV) cosmic rays are great mysteries in modern astrophysics. The current
theories for their explanation include the so-called ''top-down'' decay
scenarios whose main signature is a large ratio of UHE gamma rays to
protons. Important step in determining the primary composition at
ultra-high energies is the study of air shower development. UHE gamma ray
induced showers are affected by the Landau-Pomeranchuk-Migdal (LPM) effect
and the geomagnetic cascading process. In this work extensive simulations
have been carried out to study the characteristics of air showers from UHE
gamma rays. At energies above several times $10^{19}$ eV the shower is
affected by geomagnetic cascading rather than by the LPM effect. The
properties of the longitudinal development such as average depth of the
shower maximum or its fluctuations depend strongly on both primary energy
and incident direction. This feature may provide a possible evidence of the
UHE gamma ray presence by fluorescence detectors.
\end{abstract}

\section{Introduction}

The ultra-high energy (UHE; $>10^{19}$ eV) cosmic ray research has been
initiated about 40 years ago. Several experiments have been carried
out and at present AGASA \cite{AGASA}, HiRes \cite{HiRes} and Yakutsuk
 \cite{Yakutsk} experiments are in operation to observe air showers
initiated by UHE cosmic rays (see for a review \cite{Nagano}). Today
the number of recorded events is already big enough to convince even
strong sceptics that the cosmic ray energy spectrum extends well beyond
the theoretically expected Greisen-Zatzepin-Kuz'min (GZK) cutoff around
$5\times 10^{19}$ eV \cite{Greisen}. The origin and nature of these
particles are still unsolved questions. The problem is that it is very
difficult to extend our understanding of particle acceleration to such
extraordinary high energies and the propagation of these particles in the
cosmic microwave background (CMB) radiation restricts the distance to their
potential sources within several tens of megaparsecs (1 Mpc $=3.1\times
10^{24}$ cm).

Various models of UHE cosmic ray origin have been proposed. They are
currently a subject of very intensive discussion (see for a
review \cite{Sigl}). Models can be categorized into two basic groups of
``scenarios'': ``bottom-up'' and ``top-down''.

Conventional ``bottom-up'' scenarios look for astrophysical sources called
``Zevatrons'' (1 Z eV = $10^{21}$ eV) that can accelerate particles to
energies in excess of $10^{20}$ eV. The composition of UHE cosmic rays
is expected to be hadronic. Possible candidates include clusters of
galaxies, active galactic nuclei (AGN) radio lobes, AGN central regions,
young neutron stars, magnetars, gamma ray bursts, etc.

In the ``top-down'' scenarios UHE cosmic rays instead of being accelerated
are generated from decay of some exotic very heavy ($10^{22}$ -- $10^{25}$
 eV) {\it X-particles} that are supposed to have been formed in the early
universe. The sources of X-particles may be topological defects (cosmic
strings, cosmic necklaces, magnetic monopoles, domain
walls) \cite{Bhattacharjee} or
long-lived super-heavy relic particles \cite{Berezinsky}.

The cascade process initiated by a super-high energy neutrino ($\sim 10^{22}$
 eV) in the relic neutrino background (the so-called Z-burst model) \cite
{Weiler} is another possible scenario which is a hybrid of astrophysical
Zevatrons with new particles \cite{Olinto}.

In general, ``top-down'' and ``hybrid'' scenarios predict a rather high flux
of UHE neutrinos exceeding the observed cosmic ray flux. Gamma rays account
for a part or most of the highest energy cosmic rays above $10^{20}$ eV,
whereas nucleons would dominate at lower energies. Thus the primary
composition, especially the gamma ray content (reffered to as
gamma/proton ratio), is a
powerful discriminator between the models of the UHE cosmic ray origin. It
should be mentioned that even within conventional ''bottom-up'' models one
can expect a significant gamma/proton ratio due to the decay of neutral
pions produced in cosmic ray interactions with the 2.7 K CMB photons. Under
certain circumstances (extragalactic magnetic field strength, distance to
the sources, maximal proton energy, slope of the proton energy spectrum,
etc.), the subsequent electromagnetic cascade in the intergalactic space can
lead to a UHE gamma ray flux comparable to the observed cosmic ray flux \cite
{Halzen,Aharonian}.

Air showers initiated by UHE gamma rays have characteristic features in
comparison with ``ordinary'' hadronic showers. Two effects must be taken
into account for a study of air shower development in case of gamma ray
primaries ----- the {\it Landau-Pomeranchuk-Migdal (LPM) } effect and {\it %
cascading in the geomagnetic field}.

The influence of the LPM effect \cite{Landau,Migdal} on shower
development has been studied by many authors during the last thirty years.
The effect reduces the Bethe-Heitler (BH) cross sections for bremsstrahlung
and pair production at energies $\gtrsim 10^{19}$ eV in the atmosphere
leading to a significant elongation of the electromagnetic shower and large
fluctuations in the shower development. Generally, this effect is well
understood although there is no commonly accepted standard code taking into
account the LPM effect in electromagnetic shower modeling. It should be
pointed out that other possible mechanisms of suppression of bremsstrahlung
and pair creation processes at extremely high energies have to be more
carefully studied \cite{Klein}.

Once the electron-positron pair is produced in UHE gamma ray interaction
with the geomagnetic field away from the Earth's surface, it initiates an
electromagnetic ''cascade'' due to synchrotron radiation before entering the
atmosphere. As a result, the energy of the primary gamma ray is shared by a
bunch of lower energy secondary particles which are mainly photons and a few
electron-positron pairs. The influence of the LPM effect on subsequent
showers in the atmosphere is significantly weakened.

The history of electromagnetic cascade calculations in the geomagnetic field
is also long enough since it started with the work of McBreen and Lambert
 \cite{McBreen}. The main results of previous works
 \cite{Aharonian,Vankov,Karakula,Kasahara,Bednarek} are in a good
agreement . Recent calculations refined the previous ones revealing some
practically important features in the cascading process. For example, in
 \cite{Stanev} it is shown that the study of two major components of the
giant air showers, the size spectra and muon content, can reveal the nature
of the UHE cosmic rays if the specific dependence on the shower arrival
direction is observed. Some observables that can be extracted from the
Pierre Auger Observatory detectors (longitudinal profiles, lateral
distribution and front curvature) are discussed in \cite{Bertou}. In \cite
{Plyash} the technique of adjoint cascade equations was applied to study UHE
gamma ray shower characteristics in the geomagnetic field and in the
atmosphere emphasizing the muon component of air shower.

The aim of the present paper is to study in details the UHE gamma ray shower
characteristics that are measurable by air fluorescence detectors. It is
also important to know the difference in longitudinal shower development
between gamma ray and hadronic showers which can be used for an effective
separation between these primary species. In the following, we will discuss
the dependence of UHE gamma ray shower characteristics on the incident
direction and the possibility of detecting such showers in the future
experiments.

\section{Electromagnetic interactions in magnetic field}

About sixty years ago soon after Auger's discovery of the extensive air
showers, Pomeranchuk \cite{Pomeranchuk} estimated the maximal energy of
primary cosmic ray electrons and gamma rays that is allowed to enter the
atmosphere after interactions with geomagnetic field. According to his
calculations, the maximal electron energy $E_{c}$ due to radiation in the
geomagnetic field is a few times $10^{17}$ eV ($\sim 4\times 10^{17}$ eV for
vertically incident electrons on the geomagnetic equatorial plane). Whatever
energy greater than $E_{c}$ electrons have, they lose their energies rapidly
down to below $E_{c}$. The analogous gamma ray energy due to pair creation
in the geomagnetic field is $\sim 6\times 10^{19}$ eV.

Similar to cascading in matter, the main elementary processes leading to
particle multiplication in magnetic field are magnetic bremsstrahlung
(synchrotron radiation) and magnetic pair production. It is well known that
essentially non-zero probabilities for magnetic bremsstrahlung and pair
production require both strong field and high energies \cite{Erber}. The
relevant parameter determining the criteria for this is:
\begin{equation}
\chi =\frac{E}{mc^{2}}\frac{H_{\perp }}{H_{\rm cr}}
\end{equation}
where $E$ is the particle energy, $H_{\perp }$ is the magnetic field
strength (the component perpendicular to the particle trajectory), $m$ is
the electron mass and $H_{\rm cr}=4.41\times 10^{13}$ G.

\begin{figure}
\begin{center}
\epsfxsize=13.5cm \epsffile{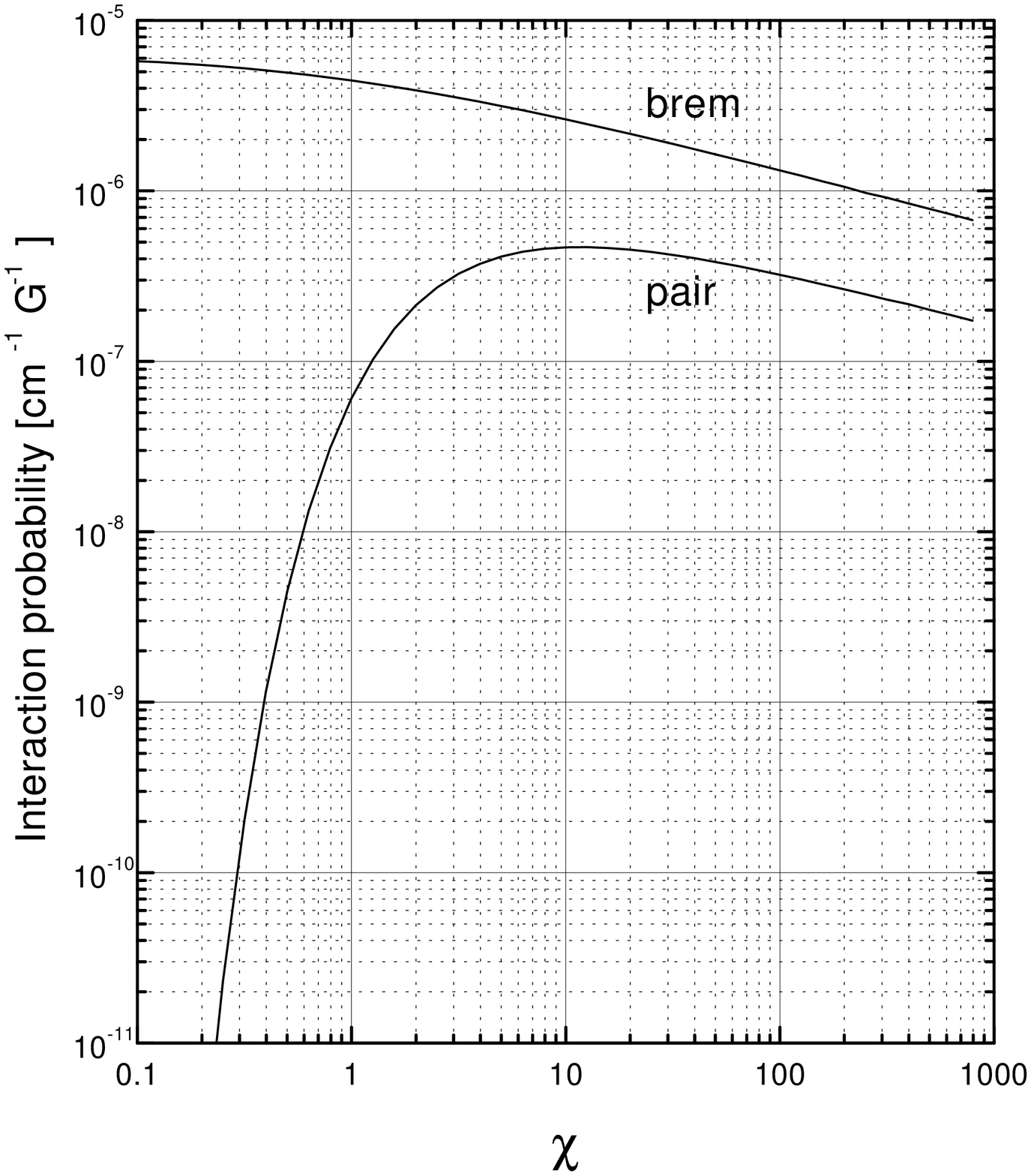}
\end{center}
\caption[]{The total probabilities (cross sections) for magnetic
bremsstrahlung and pair production as a function of parameter $\chi.$}
\label{fig:fig1}
\end{figure}

The total probabilities (cross sections) for radiation and pair production
for a given value of the magnetic field strength depend only on $\chi $ and
are shown in Fig. ~\ref{fig:fig1}. Magnetic pair production has significant probability for
$\chi \geq $ 0.1. For effective shower development, however, one needs
even higher values of $\chi \geq 1$ because the radiated photon spectrum
becomes harder with increasing $\chi $. The maximal photon energy estimated
by Pomeranchuk ($\sim 6\times 10^{19}$ eV) comes from the condition $\chi
\sim 1$.

We use the expressions of Bayer et al. \cite{Bayer} for the differential
probabilities (per unit length) for magnetic bremsstrahlung and magnetic
pair production:

\begin{equation}
\pi \left(\varepsilon ,\omega \right) d\omega =\frac{\alpha m^{2}}{\pi
\sqrt{3}}\frac{d\omega }{\varepsilon ^{2}} \left[ \left(\frac{\varepsilon -\omega }{\varepsilon }+\frac{\varepsilon }{\varepsilon -\omega }\right) K_{\frac{2}{3}}\left(\frac{
2u}{3\chi }\right) -\int\limits_{\frac{2u}{3\chi }}^{\infty }K_{\frac{1}{3}
}\left(y\right) dy\right]   \label{eq1}
\end{equation}
for bremsstralung, and
\begin{equation}
\gamma \left(\omega ,\varepsilon \right) d\varepsilon =\frac{\alpha m^{2}
}{\pi \sqrt{3}}\frac{d\varepsilon }{\omega ^{2}}
\left[ \left(\frac{\omega -\varepsilon }{\varepsilon }+
\frac{\varepsilon }{\omega -\varepsilon }\right) K_{\frac{2}{3}}\left(\frac{
2u_{1}}{3\chi }\right) +\int\limits_{\frac{2u_{1}}{3\chi }}^{\infty }K_{
\frac{1}{3}}\left(y\right) dy\right]
\end{equation}
$\mbox{for pair creation},$ where $\varepsilon $ and $\omega $ are the
electron and photon energies and $u=\omega /(\varepsilon -\omega )$, $%
u_{1}=\omega ^{2}/\varepsilon (\omega -\varepsilon )$. Here $\hbar =c=1$. $%
K_{\nu }\left(z\right) =\int\limits_{0}^{\infty }$ $e^{-z{\rm ch}(t)}%
{\rm ch}\left(\nu t\right) dt$ is the modified Bessel function known as
MacDonald's function.

While for $\chi \gg 1$ (strong field) the electromagnetic cascade develops
similar to the cascade in matter \cite{Anguelov}, in case of $\chi \leq $ 1
the photon interaction length increases sharply with decreasing photon
energy. Electrons continue to radiate and the shower becomes a bunch of
secondary photons carrying more than $94-95\%$ of the primary energy.

\section{Simulation}

In our simulation studies of air showers initiated by gamma rays and hadrons
(proton or iron), we use the AIRES code (version 2.2.1) \cite{AIRES}
incorporated with QGSJET hadronic interaction model \cite{Kalmykov}. AIRES
includes the LPM effect in simulation of electromagnetic showers. For
simulations of electromagnetic cascades in the geomagnetic field we use our
original code.

To simulate showers initiated by UHE gamma rays, we first model cascading
in the geomagnetic field starting with a single UHE gamma ray far away from
the Earth's surface down to the top of the atmosphere. Then secondary
particles that reach the top of atmosphere are set as an input for the AIRES
code. Finally, the air shower initiated by the UHE gamma ray is constructed
as a superposition of lower energy gamma ray sub-showers. In practice,
we use a library of pre-simulated showers in the atmosphere
which has been calculated by the AIRES code.

\subsection{Electromagnetic cascading in geomagnetic field}

We simulate the electromagnetic cascade in the geomagnetic field by
injecting a UHE gamma ray at a distance of $3R_{\rm e}$ away from the
Earth's surface where $R_{\rm e}$ is the Earth's radius of $6.38\times
10^{8}$ cm. The primary gamma ray and secondary particles are
propagated by taking account of pair production and synchrotron
radiation on each step (a step-size of 1km). Only particles above a
threshold energy of $10^{16}$ eV are followed in the simulation until
they reach the top of the atmosphere. This threshold energy is low
enough to neglect the contribution of sub-threshold particles in
the cascade.

In order to examine the properties of electromagnetic cascades as
a function of the incident direction, we divide uniformly the sky by
bins of $5^{\circ }$ for both azimuth and zenith angles to 1085 points
and simulate 50 events for each point. In our simulation we use the
International Geomagnetic Reference Field (IGRF) and World Magnetic
Model (WMM) which provide a good approximation for the geomagnetic field
up to 600 km above sea level \cite{IGRF}. Above this altitude the
geomagnetic field is extrapolated from this model.

In the present work we examine the properties of UHE gamma ray showers
for the location of Utah, USA (Long. = $113.0^{\circ }$ W, Lat. =
$39.5^{\circ}$N and 1500 m above sea level) near the site of the HiRes
experiment. In this location, IGRF gives the field of
0.53 G pointing $25^{\circ }$ downword from $14^{\circ }$
east of the geographical (true) north.

\subsection{Atmospheric shower simulation}

The simulations of atmospheric showers are carried out independently from
the geomagnetic cascading process. Using AIRES code, we first prepare a
library of sub-showers initiated by gamma rays with zenith angles of $%
39.7^{\circ }$, $54^{\circ }$ and $61.6^{\circ }$ and energy fixed between $%
10^{16}$ and $10^{21}$ eV at logarithmically equally spaced values (10
energies per decade). The number of all charged particles are
recorded at 5 g cm$^{-2}$ intervals in vertical depth. The library contains
500 showers at each energy. It should be noted that this number is
significantly greater than the maximal number ($\sim 100$) of secondary
gamma rays from the geomagnetic cascade in each energy bin.

The construction of UHE gamma ray initiated shower is carried out by a
method similar to the so-called ''boot-strap'' method as following: $i$-th
secondary particle with energy $E_{i}^{({\rm GM)}}$ at the top of
atmosphere is followed by a sub-shower with the nearest energy $E_{i}^{(%
{\rm atm})}$ selected randomly from the library. The secondary electron
is replaced by a gamma ray with the same energy. By summing up the
sub-showers $(i=1,\ldots ,N_{\gamma }^{({\rm GM})})$ with a weight $%
w_{i}=E_{i}^{({\rm GM})}/E_{i}^{({\rm atm})}$, the atmospheric shower
initiated by a single gamma ray with energy $E_{0}^{(\gamma )}$ is
represented as a superposition of sub-showers with total energy
${\displaystyle \sum^{N_{\gamma }}w_{i}E_{i}^{({\rm atm})}=E_{0}^{(\gamma )}}$.

In the present work we also aimulate amples of 500 hadron (proton and iron)
initiated showers for the same zenith angles and energy range to compare the
results.

\section{Result}

\subsection{Properties of geomagnetic cascading}

The cascade development is determined by the features of the cross sections
of the processes and the field strength. The maximal values of the
parameter $\chi $ which governs cascading do not much exceed 1, e.g. $\chi
=1.33$ for $10^{20}$ eV and $H_{\perp }=0.3$ G, and $\chi =13.3$ for $10^{21}
$ eV and same $H_{\perp }$, i.e. one can expect only a few gamma ray
interactions (pair creations) in the shower. But such $H_{\perp }$ values
are characteristic for the surface of the Earth. The field strength rapidly
decreases with the geocentric distance, $\sim 1/{R_{e}}$,
which means that the cascade starts not far from the surface of the Earth.
The first interaction of the gamma rays with
the energies of interest occurs in relatively narrow range of distances not
further than $3R_{e}$ . For example, the mean altitude of the first
interaction of vertical gamma rays with primary energy $E_{0}^{(\gamma
)}=10^{21}$ eV is about 5300 km.

 The typical shower profiles averaged over
1000 showers for $E_{0}^{(\gamma )}=10^{20}$ eV and different threshold
energies for secondary photons and electrons are shown in Fig. 2 (bottom
panel). The zenith angle is $40^{\circ }$ and the azimuth corresponds
to north (strong field). For comparison in this figure is also shown the
shower profile for the secondary photons with energy above $10^{16}$ eV in a
shower coming from south (weak field). This picture differs too much from
the cascade development in the matter. As the primary energy distributes
between the electron and photon components (see the top panel of Fig. 2), the
mean photon energy decreases and when $\chi <1$ the mean free path for
pair production sharply increases (see Fig. 1). The energy of the electron
component starts to return quickly to the photon component and the shower
converts to a beam of photons carrying the bulk of the primary energy.
For the case shown in Fig. 2 (strong field) the mean number of gamma ray
interactions per shower in the geomagnetic field is approximately $1$ (1.11
for the sample of 1000 showers) which means that there is almost no pair
creation after the initial interaction. For $E_{0}^{(\gamma )}=5\times
10^{20}$ eV and same conditions this number increases to 4.16. This is the
reason for the much smaller number of electrons than photons in the shower.
In our example, for the strong field, the mean number of electrons reaching
the top of the atmosphere, is 2.22 (=$2\times 1.11$) for $10^{16}$ eV
threshold energy. These electrons carry $\sim 2\%$ of the primary
energy. In the case of weak field (line with symbols in Fig. 2), the
probability of the primary gamma ray to interact in the geomagnetic field is
only $\sim 6-7\%$ and this happens on an average of $200$ km from the sea
level.

In Utah, the southern sky region is close to the direction of the
geomagnetic field and hence the effect of geomagnetic cascading is
relatively small. Gamma rays arriving from the northern sky region
 travel through stronger field whose strength increases with the zenith
angle. Generally, primary gamma rays are most affected by the geomagnetic
field when they come from the northern sky or near the horizon.

Fig. 3 shows maps for the gamma ray conversion (interaction) probability
(grey scale) with the geomagnetic field for all incident directions in
horizontal coordinates. The different panels correspond to primary
energies $E_0^{(\gamma)}=10^{19.5}$, $10^{20}$, $10^{20.5}$ and
$10^{21}$ eV. The radial coordinate is the zenith angle $\theta $. The
inner circles correspond to $\theta =30^{\circ }$ and $60^{\circ }$, and
the outer one is of the horizon. Azimuths are as labeled `N' denotes the
true north). Dashed curves indicate the opening angles of $30^{\circ }$,
$60^{\circ }$ and $90^{\circ }$ to the local magnetic field.

The region with smaller probability is around the direction which is
parallel to the local geomagnetic field. From this region, primary gamma
rays are most likely to enter the atmosphere without interaction.
Thus, this region can be referred to as ``window'' for the primary gamma
rays. Through this window they can reach the top of the atmosphere
surviving interaction and be observed as `LPM showers'. The size of this
window shrinks rapidly with increasing primary energy and almost all
gamma rays with $E_{0}^{(\gamma )}$ $\gtrsim 10^{20}$ eV initiate a
geomagnetic cascade above the atmosphere.

Fig. 4 shows maps of the average multiplicity of the secondary particles
(number of electrons plus photons above $10^{16}$ eV; grey scale) at the top
of atmosphere on the same coordinates as in Fig. 3. The average energy of
secondary particles ($E_{0}^{(\gamma )}$/multiplicity) is plotted in Fig. 5.

The patterns of these maps match well the field strength, i.e. the direction
of the geomagnetic field at the ground level, which reflects the fact that
the first interaction occurs not far from the Earth's surface.

For low primary energy and/or weak field strength, i.e. small $%
E_{0}^{(\gamma )}H_{\perp }$, the primary gamma ray does not or only once
produces electron-pair before reaching the top of the atmosphere. The
produced electrons
continuously radiate photons by magnetic bremsstrahlung and the multiplicity
increases with the primary energy. Nevertheless, radiated photons can hardly
interact again with the geomagnetic field unless $\chi $ $\gtrsim 1$.

For the direction with very strong field (or very high primary energy), i.e.
large $E_{0}^{(\gamma )}H_{\perp }$, e.g. for the northern sky region or
near-horizontally incident gamma rays, the multiplicity is almost
proportional to the primary energy which leads to a nearly constant average
particle energy at the top of the atmosphere (see Fig. 5). This energy is
about a few times of $10^{17}$ eV. In the sky regions where the conversion
probability is 100\% the maximal average energy of the secondary particles,
which are mainly photons, do not exceed several times $10^{19}$ eV. This is
consistent with the estimation by Pomeranchuk \cite{Pomeranchuk} described
previously and thus the LPM effect is ineffective in the atmosphere except
for the window regions.

The multiplicity distribution of secondary particles at the top of the
atmosphere for different incident zenith angles ($39.7^{\circ }$, $54^{\circ
}$ and $61.6^{\circ })$ and azimuths corresponding to north and south,
are plotted in Fig. 6. Each histogram includes 1000 simulated showers.

The columns in the first bin of each panel correspond to the primary gamma
rays that do not interact in the geomagnetic field. In
regions with 100\% conversion probability the fluctuations are small for
large $E_{0}^{(\gamma )}H_{\perp }$ due to the better cascade development
by the more gamma ray interactions above the astmosphere.

Fig. 7 shows the average energy distribution (spectrum) of secondary
particles at the top of the atmosphere. The data belong to the same set of
simulated showers as in Fig. 6. Here the `surviving' primary gamma rays
manifest themselves by the columns in the last bins of some histograms.

The maximum of the spectrum shifts towards the higher energies when
$E_{0}^{(\gamma )}$ increases. However, except for the cases where the
probability of survival primaries is large, i.e. the conversion
probability is not close to 100\%, this shift slows down for
$E_{0}^{(\gamma )}$ $>10^{20}$ eV and the shape of the spectrum remains
almost same for the highest energies. As mentioned previously, the
multiplicity of secondaries is almost proportional to $E_{0}^{(\gamma )}$.
The spectra display sharp cutoff at a few times $10^{19}$ eV and
subsequent sub-showers in the atmosphere are not affected by the LPM
effect. We discuss this in the next subsection.

We estimate the lateral spread of particles at the top of the atmosphere.
All particles are contained within a radius of about 10 cm. This
value is much greater than 0.1 mm given in \cite{McBreen,Kasahara}
but is still too small to be taken into account. Thus, successive shower
developments in the atmosphere can be correctly expressed as a superposition
of atmospheric sub-showers initiated by secondary particles with such energy
spectra.

\subsection{Shower development in the atmosphere}

Fig. 8 shows examples of individual shower developments initiated by
primary gamma rays with $E_0^{\gamma}=10^{19},10^{20}$ and $10^{21}$ eV.
Left and right panels correspond to azimuths from north and south,
respectively. Zenith angles are $39.7^{\circ }$, $54^{\circ }$ and $%
61.6^{\circ }$. Dashed curves represent the average shower profiles for
proton primaries.

This figure illusrates very well a remarkable feature of the shower
development at these energies --- significantly small fluctuations in the
case that the primary gamma ray interacts with the geomagnetic field. The
largest fluctuations in the shower development can be found for
$E_0^{\gamma}=10^{20}$ eV, $\theta =39.7^{\circ }$ and incident direction
from south. In this case the primary gamma ray conversion probability in
the geomagnetic field is only several percent and the LPM effect affects
strongly the shower development in the atmosphere. Increasing primary
energy leads to significant decrease of fluctuations and this trend is
stronger for the sky regions close to the horizon (large $H_{\perp }$).
These two effects are more pronounced for northern sky regions.

Fig. 9 shows distributions of the average depth of the shower maximum in
the atmosphere $\langle X_{{\rm max}}\rangle $ for showers with
$E_0^{(\gamma)}=10^{19}$, $10^{19.5}$, $10^{20}$, $10^{20.5}$ and $10^{21}$
 eV. We present the cases of zenith angles of $54^{\circ }$ and
$61.6^{\circ }$ in left and right panels, respectively. In each panel, the
solid line indicates the distribution of proton showers. Dotted and dashed
lines represent those for gamma ray showers from azimuths of north and
south, respectively.

The shape of $\langle X_{{\rm max}}\rangle $ distributions of gamma ray
showers noticeably varies with primary energy and incident direction. As the
primary energy increases, the distributions become narrower and the mean
values are almost constant for energies $>10^{20}$ eV (see also Fig. 11).
Some distributions having two maxima or long tail to the deeper
$X_{{\rm max}}$ result from a mixture of converted and not-converted
gamma rays above the atmosphere.

In Fig. 10, maps of $X_{{\rm max}}$ are shown on the same coordinates as
in Figs. 3 -- 5.

The data for this figure are obtained by the following approximation. The
simulation of the shower development is based on the method described in the
previous section. For a given zenith angle we use data from the library of
pre-simulated showers with similar zenith angle. Since the magnitude of the
LPM effect depends on the atmospheric density profile along the particle
trajectory, there is a small dependence on the incident zenith angle. In the
vertical region for $10^{20}$ eV, this effect possibly matters and
$\langle X_{{\rm max}}\rangle $ can be higher than shown due to the lack of
air shower library for this direction. For higher primary energy or other
sky region, however, primary gamma rays are converted and thus it is not
necessary to take into account the LPM effect for the estimation of
$\langle X_{{\rm max}}\rangle $.

In each map, i.e. for each $E_{0}^{(\gamma )}$, $\langle X_{{\rm max}%
}\rangle $ reflects the dependence of geomagnetic cascading on the incident
direction. Typically, gamma ray showers with larger $\langle X_{{\rm max}%
}\rangle $ are predominantly coming from southern sky region. Only for $%
10^{20}$ eV there is a small window where gamma ray showers are affected by
the LPM effect. This region may serve as a probe for UHE gamma ray presence.

Fig. 11 shows the relation between $\langle X_{{\rm max}}\rangle $ and $%
E_{0}^{(\gamma )}$ for gamma ray showers. For comparison, corresponding
relations for
proton and iron primaries are also drawn in the figure. Incident azimuths
of gamma rays are from north and south. Dashed lines and thick solid lines
are for zenith angles of $54^{\circ }$ and $61.6^{\circ }$, respectively.
The dotted curve indicates the case of no geomagnetic field.

For proton and iron showers $\langle X_{{\rm max}}\rangle $ increases
with $E_{0}^{(\gamma )}$ and the slope of the relation, i.e. the elongation
rates are almost constant and are 54 and 56 g cm$^{-2}$/decade,
respectively.
The elongation rate for gamma ray showers is greater than those of hadronic
ones and is also constant according to the electromagnetic cascade theory up
to energies $\sim 2\times 10^{19}$ eV. Above this energy the LPM effect
steepens the relation of $\langle X_{{\rm max}}\rangle $ versus $%
E_{0}^{(\gamma )}$ as shown by the dotted line in the figure.

The geomagnetic cascading starts to ``work'' approximately at the same
energy, about several times $10^{19}$ eV. At this energy which depends on the
incident direction, $\langle X_{{\rm max}}\rangle $ reaches its maximum.
Above this energy the geomagnetic cascade develops well enough to suppress
the LPM effect in the atmosphere. This leads to a rapid decrease of $\langle
X_{{\rm max}}\rangle $.

The slow increase of $\langle X_{{\rm max}}\rangle $ after its minimum
results from the slow increase of the fraction of secondary photons above
the threshold for the LPM effect (see Fig. 7). The multiplicity also
increases proportionally to $E_{0}^{(\gamma )}$ which leads to almost
constant average photon energy in the bunch above $10^{20}$ eV. It must be
noted that a superposition of BH sub-showers has smaller $\langle
X_{{\rm max}}\rangle $ than a single BH shower with an energy equal to
the sum of the sub-shower energies.

In Fig. 12, the fluctuations of $X_{{\rm max}}$ (the standard deviation of
$X_{\max }$ distribution) are shown as a function of primary energy .
The line key is the same as in Fig. 11 but the case of no geomagnetic
field is not shown.

Similar to $\langle X_{{\rm max}}\rangle $, the fluctuations for
gamma ray showers vary typically depending on primary energy and incident
direction, while those for proton and iron showers are almost constant, $%
\sim 67$ g cm$^{-2}$ and $\sim 26$ g cm$^{-2}$, respectively.

For gamma ray showers, the picture between $19^{19}$ and $10^{20}$ eV is
similar to that in Fig. 11. Large fluctuations are due to the LPM effect.
For energies at which the geomagnetic cascading is effective, the
fluctuations decrease rapidly with the energy. At the highest energies,
the fluctuations tend to be as small as those for iron showers. This
behavior attributes to a competition between the LPM effect and
geomagnetic cascading as in Fig. 11.

\section{Discussion and conclusions}

Our study shows that the longitudinal development of gamma ray showers is
not simply scaled with primary energy. Shower development in the energy
region above $\sim 10^{19.5}$ eV shows very specific dependence both on the
primary energy and incident direction. $\langle X_{{\rm max}}\rangle $ of
gamma ray showers is larger than that expected for proton showers.
Furthermore, the elongation rate of gamma ray showers shows considerable
variation with energy depending on their incident directions. The future
observation of these longitudinal shower characteristics with better
statistical accuracy would be the possible key for studying the UHE
gamma ray flux. Also, additional information may be obtained from the
properties of $X_{{\rm max}}$ fluctuations.

In order to acquire a definite conclusions of the primary composition of UHE
cosmic rays, more elaborate considerations may be required due to a limited
statistics and difficulty in separating between gamma ray and hadronic
showers. For example as is shown above, the fluctuations of $X_{{\rm max}}$ for
gamma ray showers become even less than those for proton showers and are
 close to
iron showers above $10^{20}$ eV. We expect that the development of a number
of UHE showers would be measured with a better accuracy in the near future
which will be the first decisive step in looking for UHE gamma rays.

If the observed showers develop slowly from the sky region nearby the
window (see Figs. 2 -- 4 and 9) showing typical characteristics of LPM
showers, this could be a noticeable and physically important evidence of \
the UHE gamma ray presence.

As was earlier mentioned in \cite{Stanev}, magnetic bremsstrahlung process
may be important for the shower development at high altitudes where the
atmospheric density is very low. According to the estimations in \cite
{Goncharov}, made by numerical integration of the system of cascade
equations, the interaction of shower particles with the geomagnetic field
inside the atmosphere becomes important for cascades created by primary
gamma rays with energies higher than $\sim 3\times 10^{20}$ eV. For example,
injecting the gamma rays with energy $10^{20}$ eV vertically into the
atmosphere, the number of particles in showers at sea level calculated only
with LPM effect is $\sim 13\%$ less than that in showers when interactions
with the geomagnetic effect (for $H_{\perp }=0.35$ G) taken into account.
This difference increases with the primary energy up to $\sim $ 2.5
times for $10^{21}$ eV. Using our own simple hybrid code for one-dimensional
atmospheric shower simulation we obtain similar results showing also a
noticeable shift of the shower maximum. This work is now in progress.

Planned projects (Auger \cite{Auger}, EUSO \cite{EUSO}, etc.) to study UHE
cosmic rays promise to observe individual shower development with better
accuracy. It is important to find the effective and reasonable physical
parameters from simulation studies in order to discriminate gamma ray
showers from hadronic ones on event by event basis. It can be also strong
probe by including the fluctuation study in addition to average shower
development in the discussions about the composition of UHE cosmic rays.

\acknowledgments

We thank T. Stanev for valuable help and discussions. H.P.Vankov is thankful
to Japan Society for the Promotion of Science(JSPS) for support of his visit
to Japan where this work was conceived, and to the National Graduate
Institute for Policy Study (GRIPS) for its hospitality.

\begin{figure}[p]
\begin{center}
\epsfxsize=13.5cm \epsffile{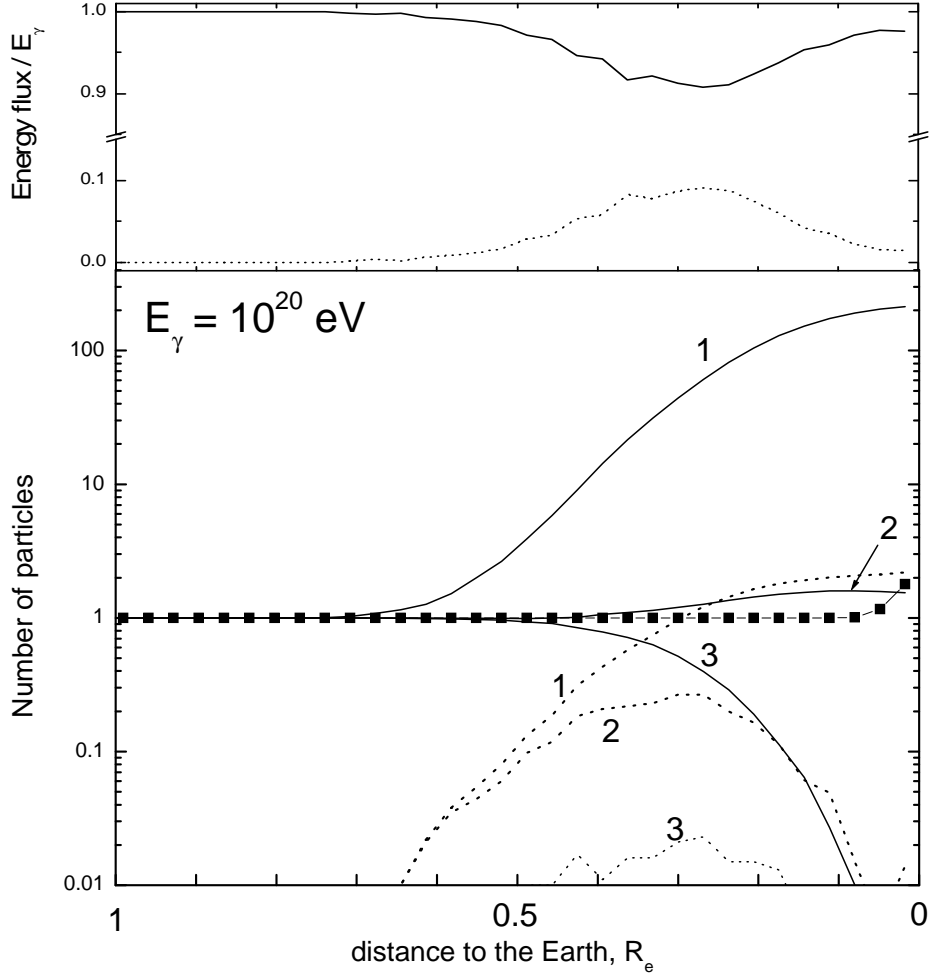}
\end{center}
\caption[]{Shower profile (bottom panel) in the geomagnetif field for
primary gamma ray with energy $10^{20}$ eV and different threshold
energies: 1---$10^{16}$ eV, 2---$10^{19}$ eV and 3---$5\times 10^{19}$
 eV. The zenith angle is 40$^\circ$ and the azimuths corresponds
to the north. Solid and dotted curves indicates for photons and
electrons, respectively. Curves with symbols show the number of
photons with energies $>10^{16}$ eV in showers with the same primary
energy and azimuth from the south. The energy flux carried by photons
(solid curve) and electrons (dotted curve) is shown on the top panel
for azimuth from the north. $R_{\rm e}$ is the Earth's rudius$(=6.38\times
10^{8}{\rm cm})$.}
\end{figure}

\begin{figure*}
\begin{center}
\epsfxsize=13.5cm \epsffile{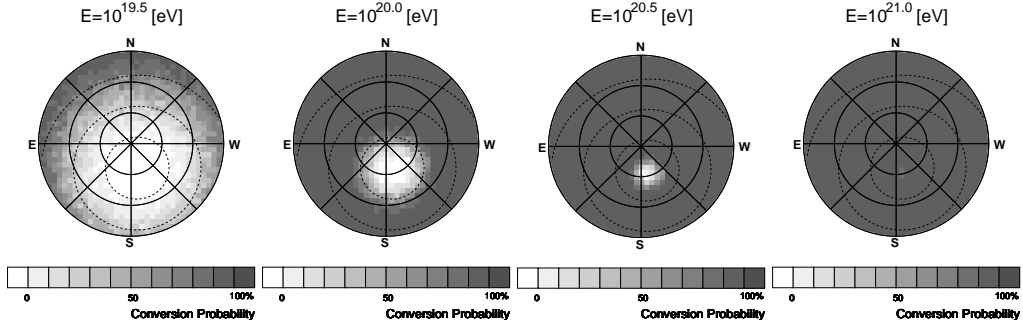}
\end{center}
\caption{Maps of gamma ray conversion probability in the geomagnetic field
for primary energies $10^{19.5}$, $10^{20}$, $10^{20.5}$ and $10^{21}$ eV.
Inner circles correspond to zenith angles $30^{\circ }$, $60^{\circ }$ and
horizon.}
\end{figure*}

\begin{figure*}
\begin{center}
\epsfxsize=13.5cm \epsffile{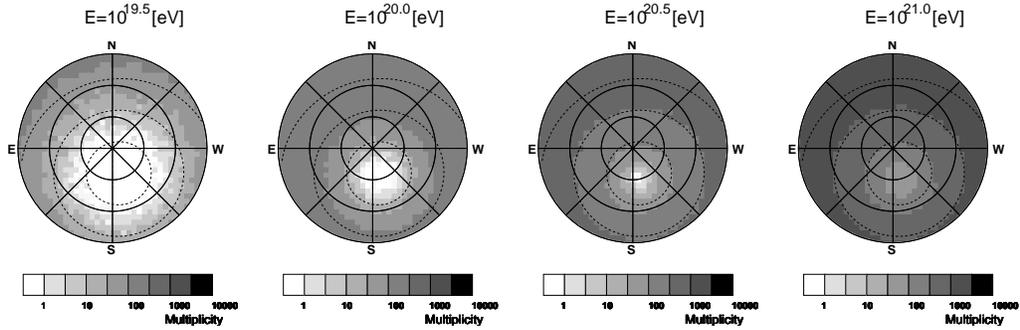}
\end{center}
\caption{Maps of average multiplicity of secondary particles with energy $%
>10^{16}$ eV at the top of atmosphere. Primary gamma ray energies and
coordinates are the same as in Fig. 3.}
\end{figure*}

\begin{figure*}
\begin{center}
\epsfxsize=13.5cm \epsffile{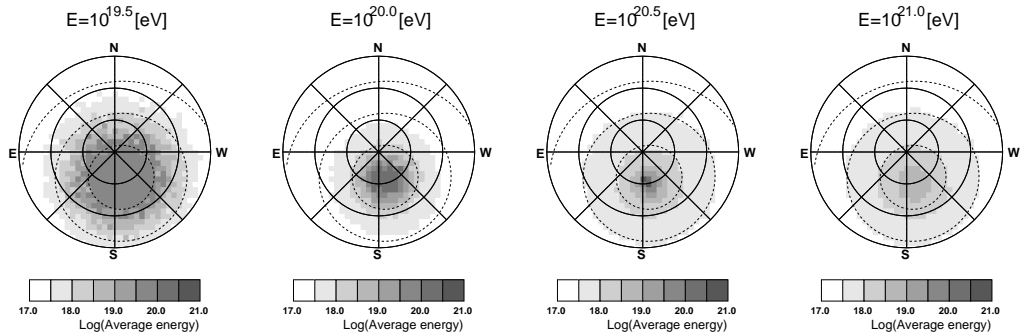}
\end{center}
\caption{Maps of average energy of secondary particles with energy
$>10^{16}$ eV at the top of atmosphere. Primary gamma ray energies and
coordinates are the same as in Figs. 3 and 4.}
\end{figure*}

\begin{figure}
\begin{center}
\epsfxsize=13.5cm \epsffile{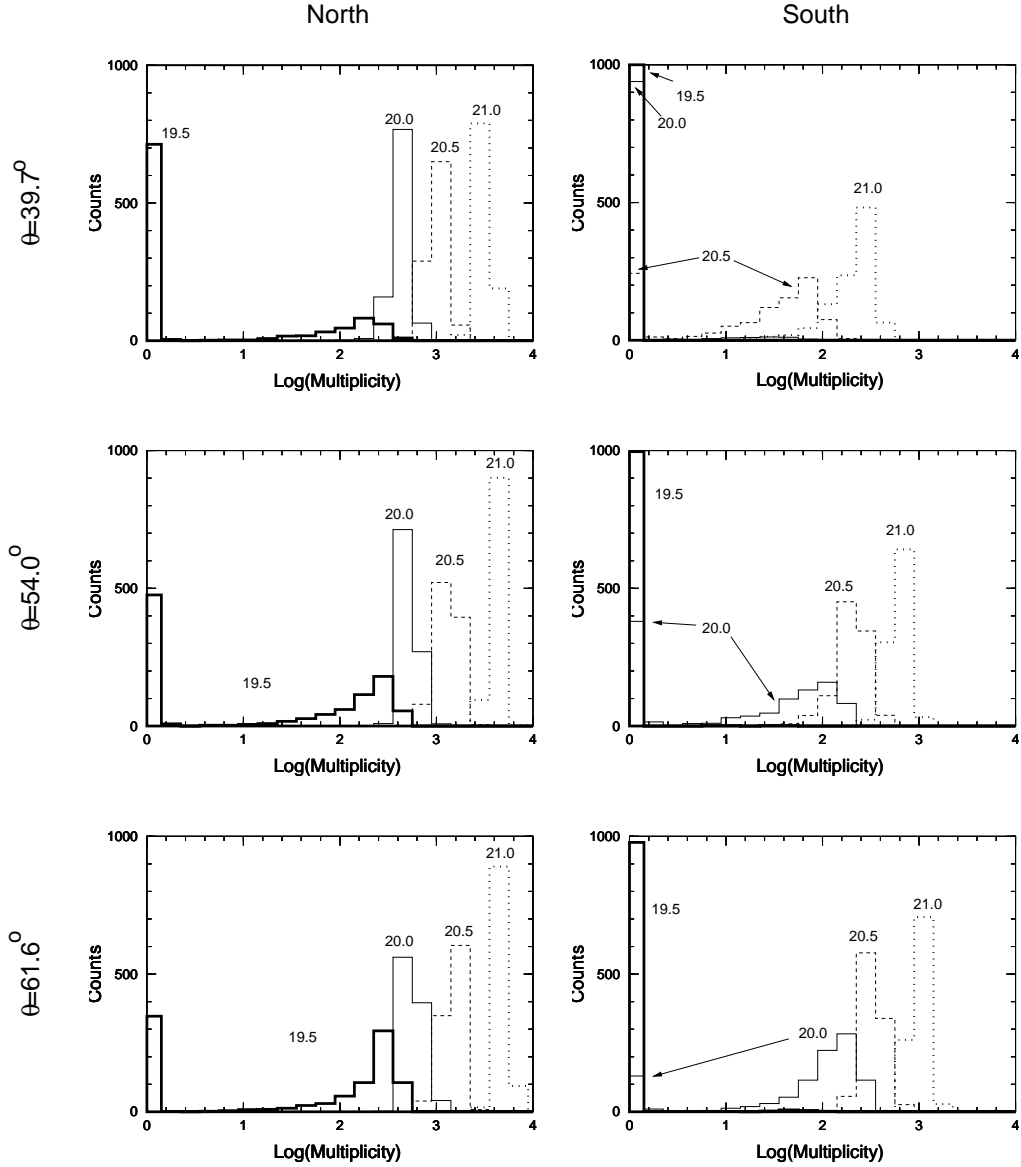}
\end{center}
\caption[]{Multiplicity distribution of secondary particles (photons plus
electrons) at the top of atmosphere for primary energies of $10^{19.5}$, $
10^{20}$, $10^{20.5}$ and $10^{21}$ eV and different zenith angles of
$39.7^{\circ }$, $54^{\circ }$ and $61.6^{\circ }$. Arrival directions of
gamma rays are from north and south.}
\label{fig:mult}
\end{figure}

\begin{figure}
\begin{center}
\epsfxsize=13.5cm \epsffile{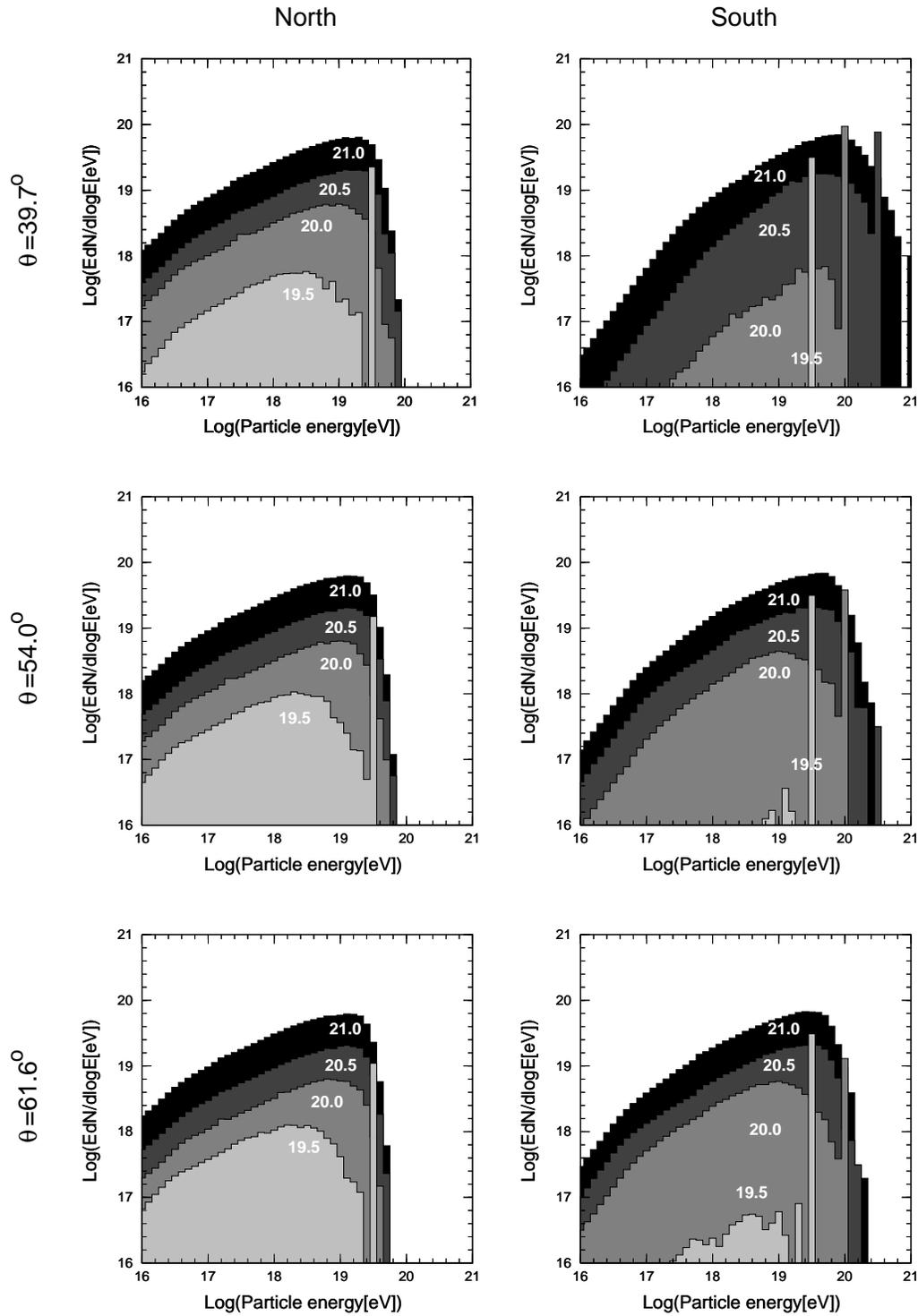}
\end{center}
\caption[]{Energy distribution (spectrum) of secondary particles
(photons plus electrons) at the top of atmosphere. Each panel corresponds
to that in Fig. ~\ref{fig:mult}.}
\label{fig:spec}
\end{figure}

\begin{figure}
\begin{center}
\epsfxsize=13.5cm \epsffile{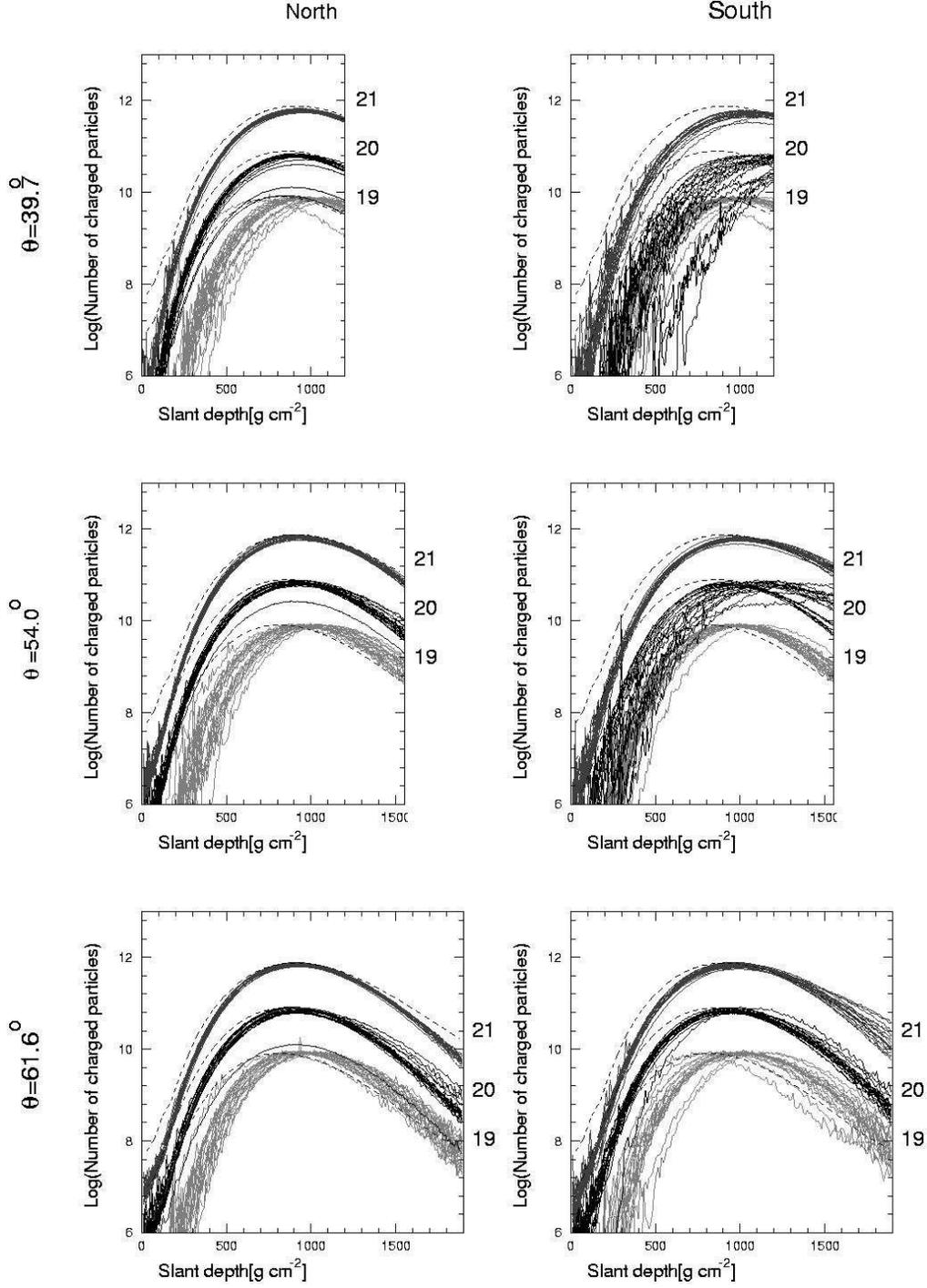}
\end{center}
\caption[]{Longitudinal development of individual gamma ray showers in the
atmosphere for primary energies of $10^{21}$, $10^{20}$ and $10^{19}$ eV
(from top) and different zenith angles of $39.7^{\circ }$, $54^{\circ }$
and $61.6^{\circ }$. Arrival azimuths are from true north and south. Dashed
curves correspond to average shower developments for proton primaries
calculated with QGSJET model.}
\label{fig:longs}
\end{figure}

\begin{figure}
\begin{center}
\epsfxsize=13.5cm \epsffile{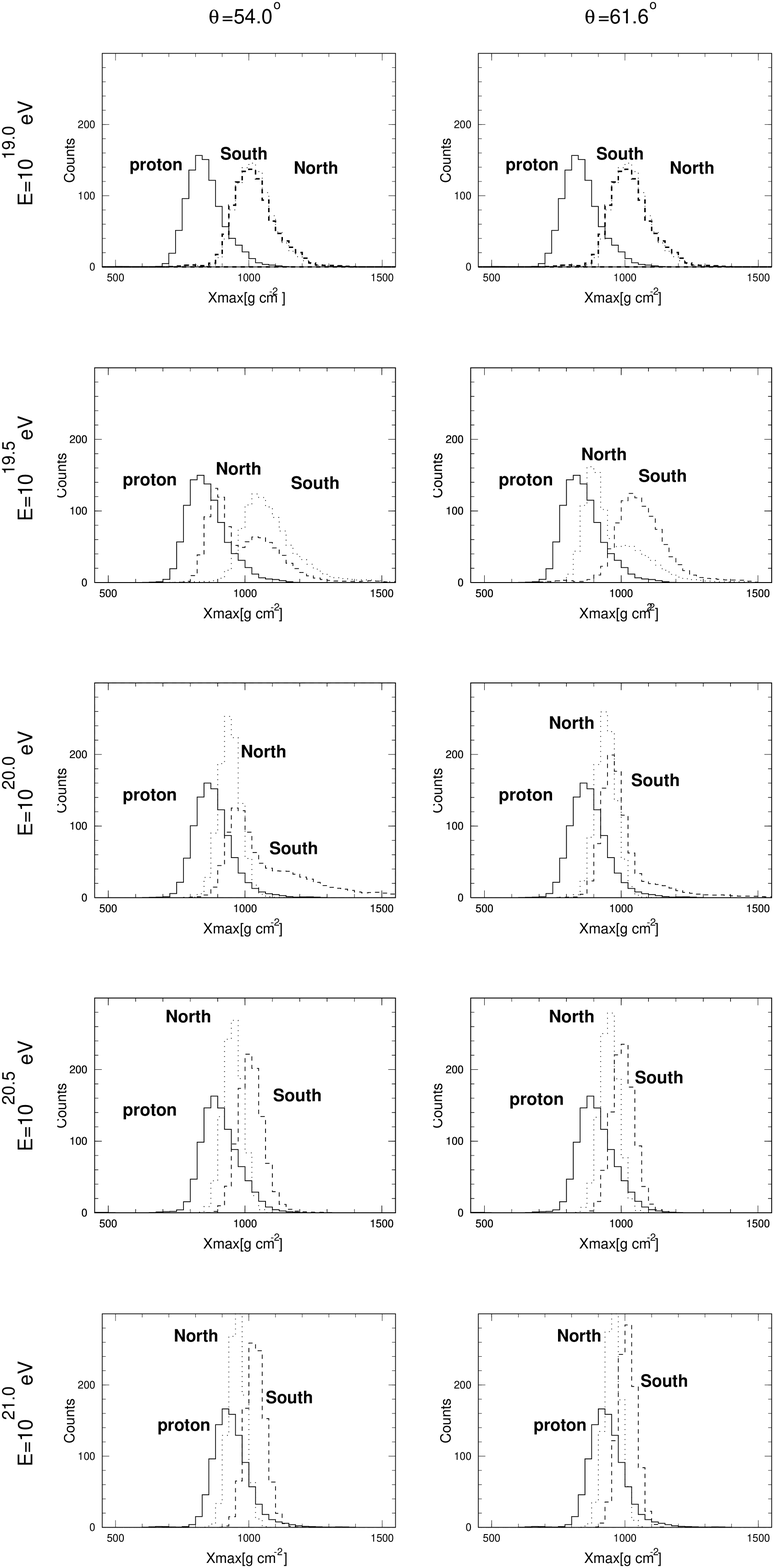}
\end{center}
\caption[]{$X_{\rm max }$ distributions for proton and gamma ray showers for
primary energies of $10^{19}$, $10^{19.5}$, $10^{20}$, $10^{20.5}$ and
$10^{21}$ eV and different zenith angles of $39.7^{\circ }$, $54^{\circ }$
and $61.6^{\circ }$. Azimuths are north (dotted lines) and south (dashed
lines).}
\label{fig:max-dist}
\end{figure}

\begin{figure*}[t]
\begin{center}
\epsfxsize=13.5cm \epsffile{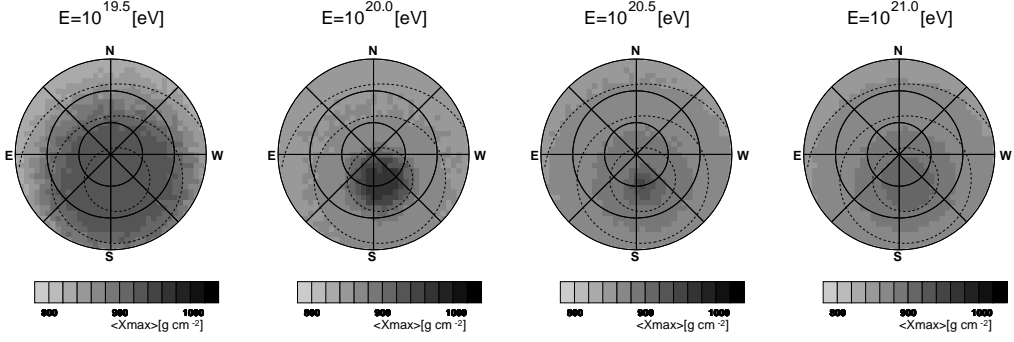}
\end{center}
\caption{Maps of average depth of shower maximum $\langle X_{\rm max }\rangle $
in the atmosphere for primary energies $10^{19.5}$, $10^{20}$, $10^{20.5}$
and $10^{21}$ eV. Coordinates are the same as in Fig. 4.}
\label{fig:ave_max}
\end{figure*}

\begin{figure}
\begin{center}
\epsfxsize=13.5cm \epsffile{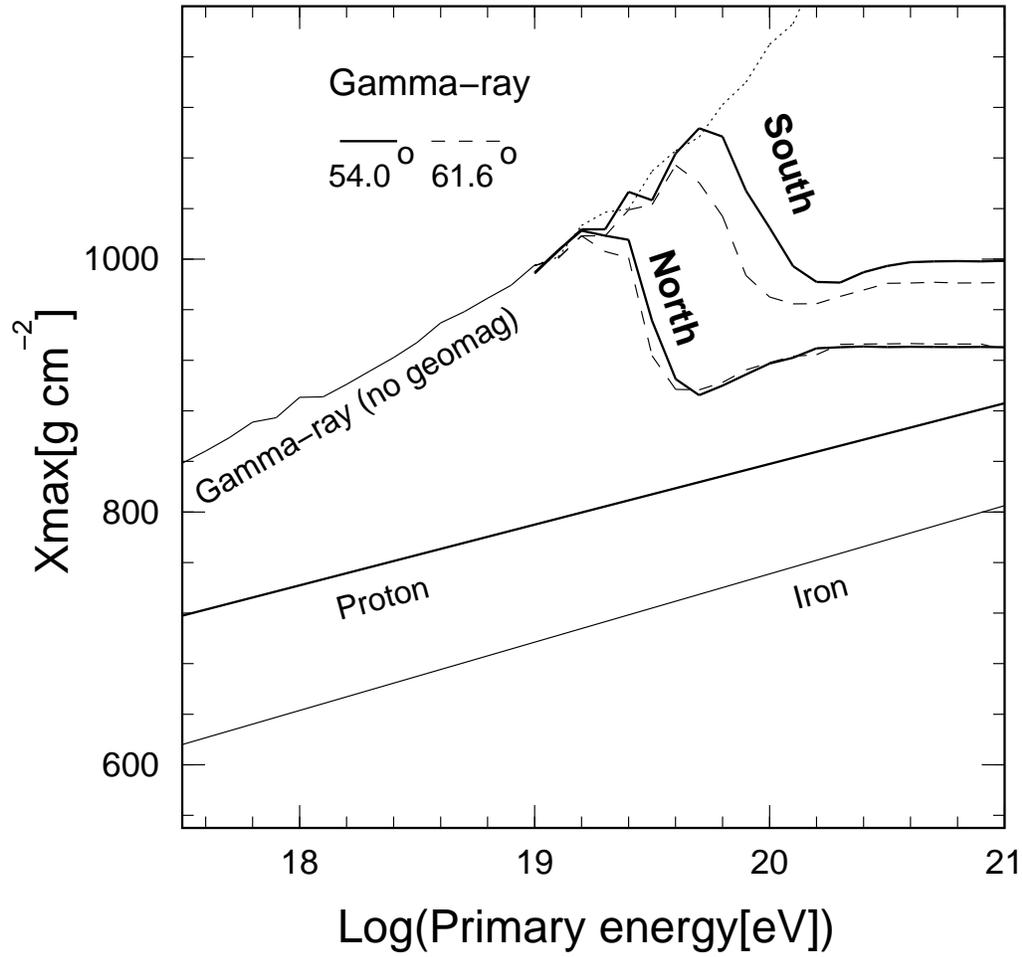}
\end{center}
\caption[]{The average depth of shower maximum
$\langle X_{\rm max }\rangle$
in the atmosphere as a function of primary energy
for gamma ray showers. Corresponding relations for proton and iron are
also drawn by solid lines. Arrival directions of gamma rays are from
north and south as denoted. The dotted curve indicates the case of no
geomagnetic field. Dashed line and thick solid curves are for zenith
angles of $54^{\circ }$ and $61.6^{\circ }$, respectively. }
\label{fig:pen-max}
\end{figure}

\begin{figure}
\begin{center}
\epsfxsize=13.5cm \epsffile{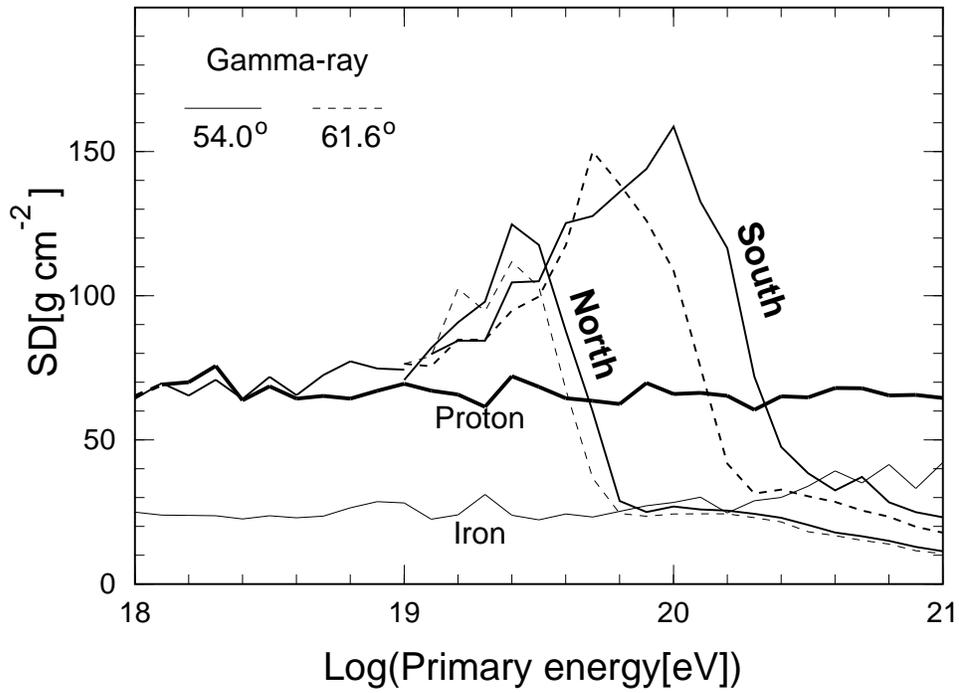}
\end{center}
\caption[]{Fluctuations (standard deviation $\sigma $) of $X_{\rm max }$ as a
function of primary energy. The line key is as in Fig. ~\ref{fig:pen-max} but
the case of no geomagnetic field is not shown.}
\label{fig:dev}
\end{figure}

\end{document}